\documentstyle[aps,eqsecnum,preprint,epsf]{revtex}
\tightenlines
\begin{document}
\draft
\preprint{}
\title{Measuring the finite width and unitarity corrections to
the $\phi\omega$ mixing amplitude.}
\author{N.N. Achasov and A.A. Kozhevnikov}
\address{Laboratory of Theoretical Physics, \\
Institute for Mathematics,  \\
630090, Novosibirsk-90, Russian Federation}
\date{\today}
\maketitle
\begin{abstract}
It is shown that the phase of $\phi\omega$ interference in the reaction
$e^+e^-\to\pi^+\pi^-\pi^0$ at energies close to the $\phi(1020)$ peak
can be calculated  in a way that is practically independent of the
model of $\phi\omega$ mixing. The magnitude of the
presently measured interference phase, still of poor accuracy, is in
agreement with the predictions  based on extending the $\omega(782)$
resonance tail from the peak position to the $\phi$ mass
upon assuming the $\omega\to\rho\pi\to3\pi$ model.
The calculated $\omega$ width at the $\phi$ mass is about 200 MeV.
\end{abstract}
\narrowtext

\section{Introduction}
\label{sec1}

Recent measurements of the $e^+e^-\to\pi^+\pi^-\pi^0$ reaction cross section
at energies in the vicinity of the $\phi(1020)$ resonance
reached by the CMD-2 team in Novosibirsk have revealed
the $\phi\omega$ interference phase
$\chi_{\phi\omega}=162^\circ\pm17^\circ$ \cite{cmd2}, provided the phases
of the complex propagators of  $\phi$ and $\omega$ mesons are properly
included:
\begin{eqnarray}
\sigma_{3\pi}&\propto&\left|{1\over m^2_\omega-s-i\sqrt{s}\Gamma_\omega(s)}
\right.\nonumber\\
&&\left.+{A\exp(i\chi_{\phi\omega})\over
m^2_\phi-s-i\sqrt{s}\Gamma_\phi(s)}\right|^2;
\label{expcro}
\end{eqnarray}
$A$ being a real positive number, and $A_{\rm bg}$ denoting the contribution
of the nonresonant background.
Hereafter $s$ is the total center-of-mass energy
squared.  The accuracy of the measurements is expected to be drastically
improved by the Novosibirsk SND and CMD-2 teams at the VEPP-2M facility,
not to mention the DA$\Phi$NE machine,
with its huge number of expected $\phi$ mesons.
The measured phase is still consistent (within 1$\sigma$) with the
canonical value of $180^\circ$ predicted
in approaches based on the flavor SU(3) and the simplest quark model with
real coupling constants \cite{ach75}. The canonical
phase explains correctly the
location of the $\phi\omega$ interference minimum in the energy behavior
of the $e^+e^-\to\pi^+\pi^-\pi^0$ reaction cross section above the $\phi$
mass, as observed in experiment \cite{cmd2,dol}.
However, the deviation of the central value of the measured
$\chi_{\phi\omega}$ from $180^\circ$ points, possibly, to some dynamical
source. The aim of the present work is to reveal the latter. To this end
we will demonstrate that $\chi_{\phi\omega}$ can be calculated
in a way that is practically independent of the specific model of
$\phi\omega$ mixing. As  will become clear, this is due to the
compensation between the $\rho\pi$ state contribution
to the $\phi\omega$ mixing amplitude, and the direct transition.
The deviation of $\chi_{\phi\omega}$
from $180^\circ$ will be shown to be explained mainly by the finite width
effects. The precise measurement of this
phase  could offer the firm ground for the extension of the $\omega$
excitation curve to the energies up to the $\phi$ mass.

Below, in sec.~\ref{sec2}, the basic  models of the decay $\phi\to\rho\pi$
are outlined. Section \ref{sec3} is devoted to the discussion of the
unitarity corrections to the coupling constants and the $\phi\omega$ mixing
amplitude. The $\phi\omega$ interference phase $\chi_{\phi\omega}$ is
calculated in sec. \ref{sec4}. Section \ref{sec5} contains conclusion
drawn from the work.

\section{Basic sources  of the $\phi\to\rho\pi$ decay}
\label{sec2}

\begin{figure}
\centerline{\epsfysize=5.5in \epsfbox{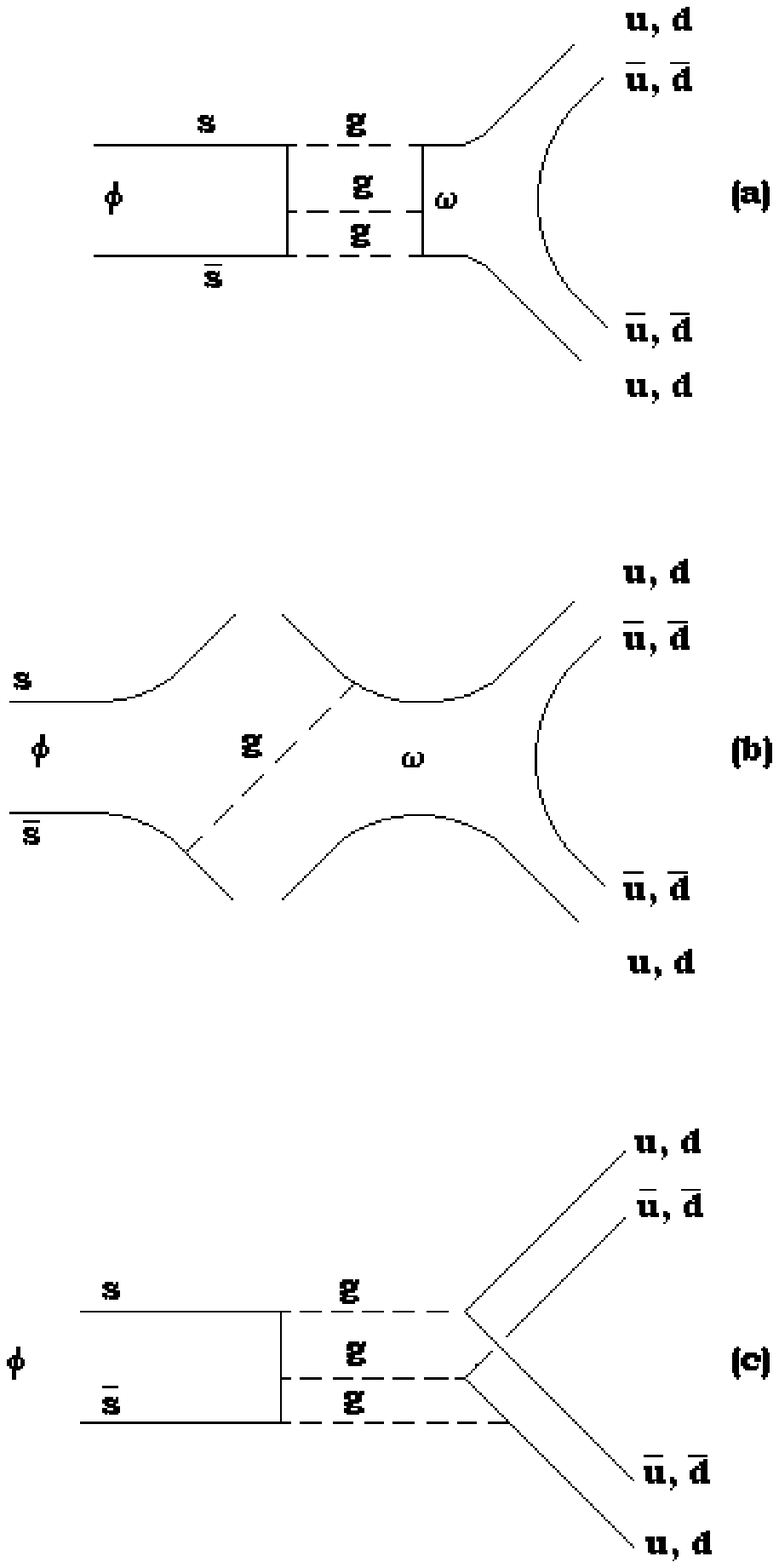}}
\caption{Possible mechanisms of the decay $\phi(1020)\to\rho(770)\pi$.
(a) The $\phi\omega$ mixing caused by the three-gluon mechanism.
(b) The $\phi\omega$ mixing due to the nonperturbative QCD effects.
Shaded regions denote the quark condensates. (c) The three-gluon mechanism
of the direct transition $\phi\to\rho\pi$.
Gluon is denoted by $g$.}
\label{fig1}
\end{figure}

All the necessary theoretical background for analyzing the $\phi\omega$
interference pattern in the cross section
of the reaction $e^+e^-\to\pi^+\pi^-\pi^0$ was developed earlier
\cite{ach92,ach93,ach95}, so one may find the  details in these
papers.
The problem of to what extent the $\omega(782)$ and $\phi(1020)$ mesons
are ideally mixed states,
\begin{eqnarray}
\omega^{(0)}&=&(u\bar u+d\bar d)/\sqrt{2},     \nonumber\\
\phi^{(0)}&=&s\bar s,
\label{unmixed}
\end{eqnarray}
is as old as these mesons themselves \cite{okubo}. The fact is that the decay
$\phi\to\rho\pi\to\pi^+\pi^-\pi^0$ which violates the Okubo-Zweig-Iizuka
(OZI) rule \cite{okubo,zweig,iizuka} is usually considered as  evidence
in favor of an admixture of the
nonstrange quarks in the wave function of  $\phi$ meson:
\begin{equation}
\phi(1020)=s\bar s+\varepsilon_{\phi\omega}(s)(u\bar u+d\bar d)/\sqrt{2},
\label{phi}
\end{equation}
where the $\phi\omega$ mixing amplitude is described by
the complex mixing parameter
$\varepsilon_{\phi\omega}(s)$ dependent on energy,
$|\varepsilon_{\phi\omega}(s)|\ll 1$. It can be
expressed through the nondiagonal polarization operator
$\Pi_{\phi\omega}$ according to the relation
\begin{equation}
\varepsilon_{\phi\omega}(s)=-\frac{\mbox{Re}\Pi_{\phi\omega}+i
\mbox{Im}\Pi_{\phi\omega}(s)}{\Delta M^2_{\phi\omega}(s)},
\label{eps}                                              
\end{equation}
where
\begin{equation}
\Delta M^2_{\phi\omega}(s)=
\Delta m^{(0)2}_{\phi\omega}-i
\sqrt{s}\left[\Gamma^{(0)}_\phi(s)-\Gamma^{(0)}_\omega(s)\right],
\label{delm}
\end{equation}
and $\Delta m^{(0)2}_{\phi\omega}=m^{(0)2}_\phi-m^{(0)2}_\omega$.
Hereafter $m^{(0)}_V\mbox{, }\Gamma^{(0)}_V$ are, respectively, the mass and
width of the ideally mixed states in
Eq.~(\ref{unmixed}), and all quantities
with the superscript (0) refer to these states. Below we will call this
mechanism the model of strong $\phi\omega$ mixing.
In QCD, the real part of the mixing operator Re$\Pi_{\phi\omega}$ arises
qualitatively
either via the perturbative three-gluon intermediate
state shown in Fig.~\ref{fig1}(a) \cite{arafune,geshken}
or the nonperturbative
effects \cite{svz} diagrammatically shown in Fig.~\ref{fig1}(b).
Quantitatively, the contribution of Fig.~\ref{fig1}(a) is small and of the
wrong sign \cite{arafune,geshken} while the calculations of
$\varepsilon_{\phi\omega}(m^2_\phi)$
according to Fig.~\ref{fig1}(b) \cite{svz} can
be considered  as order-by-magnitude estimates at best.
The one photon
contribution to Re$\Pi_{\phi\omega}$ is by two orders of magnitude smaller
than the value necessary to explain the $3\pi$ branching ratio of the $\phi$.
The non-one-photon contribution to Re$\Pi_{\phi\omega}$ is assumed
to be independent on energy. As it was pointed out in Ref. \cite{ach92},
this assumption does not contradict the data.

An alternative to
the conventional $\phi\omega$ mixing is the direct decay,
Re$g^{(0)}_{\phi\rho\pi}\not=0$,  Re$\Pi_{\phi\omega}\equiv0$
diagrammatically
shown in Fig.~\ref{fig1}(c). It is essentially the famous
Appelquist-Politzer
mechanism \cite{appel} of the OZI rule violation in the decays of
heavy quarkonia into the light hadrons, extrapolated to the $\phi$ mass
region.
As is shown in \cite{ach95}, the direct decay can be considered as
a viable contribution to
the $\phi\to\rho\pi$ amplitude \cite{fn2}. An order-of-magnitude
estimate of Re$g^{(0)}_{\phi\rho\pi}$ \cite{ach95} is in agreement with the
value extracted from the $\phi\to3\pi$ branching ratio.
This model will be called the model of weak $\phi\omega$ mixing.
Intermediate variants are possible, of course.

\section{Unitarity corrections to couplings and $\phi\omega$ mixing
amplitude}
\label{sec3}

\begin{figure}
\centerline{\epsfysize=5.5in \epsfbox{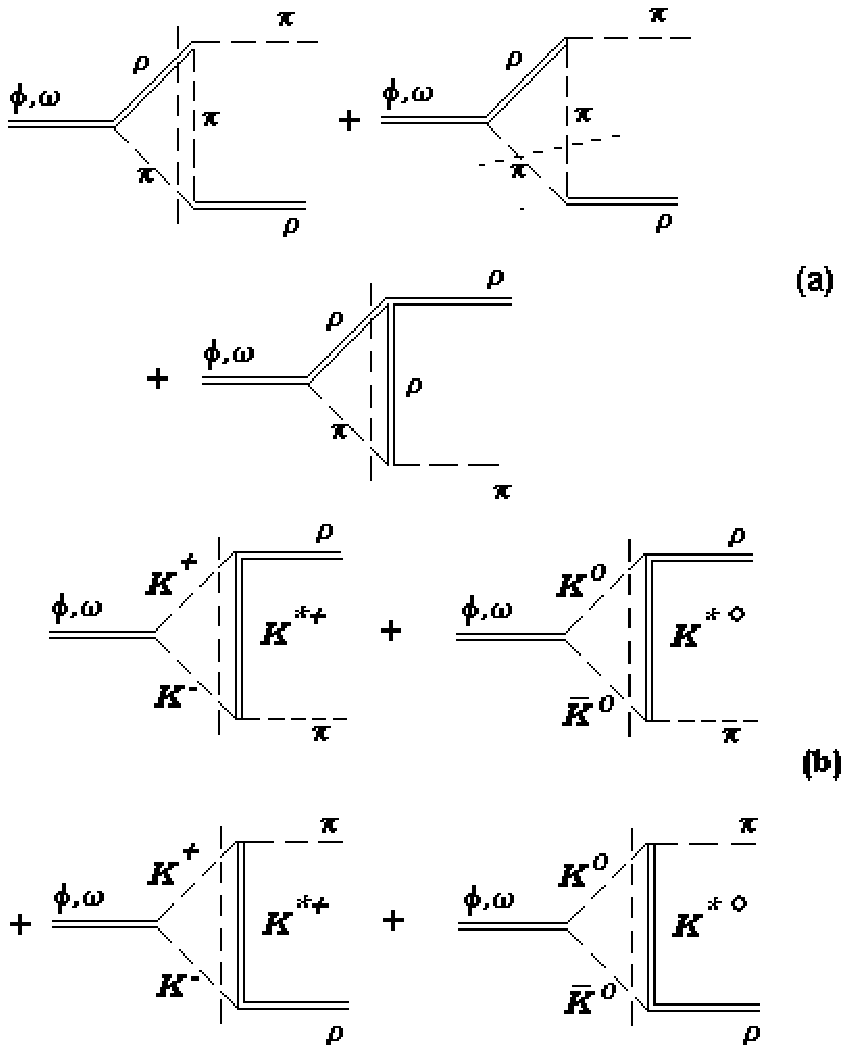}}
\caption{The contributions to 2Im$g^{(0)}_{\phi(\omega)\rho\pi}$ from
the $\rho\pi$ intermediate state (a) and
the $K\bar K$ intermediate state (b).}
\label{fig2}
\end{figure}

Contrary to Re$g^{(0)}_{\phi\rho\pi}$ and Re$\Pi_{\phi\omega}$, which are in
fact unknown, their imaginary counterparts can be evaluated reliably via
the unitarity relation.
The dominant contributions to 2Im$g^{(0)}_{\phi(\omega)\rho\pi}$
come from the diagrams shown in Fig.~\ref{fig2}.
The sum of the first two diagrams, upon
extending the results of works \cite{ach94a,ach94b} to include the form
factor of the $\pi$ exchange, $\exp(-\lambda_\pi|t-m^2_\pi|)$, is
\begin{eqnarray}
\Phi_{\rho\pi}(s,m^2)
&=&-\frac{g^2_{\rho\pi\pi}}{8\pi\sqrt{s}q_f^3}
\int_{2m_\pi}^{\sqrt{s}-m_\pi}d\mu
\frac{2\mu^2\Gamma(\rho\to\pi\pi,\mu)}{\pi|D_\rho(\mu^2)|^2}
\nonumber\\
&&\times\left\{(q_iq_f)^2\mbox{vp}\int_{-1}^{+1}dx\frac{1-x^2}{a+x}
\right.      \nonumber\\
& &\left.\times
\left[\exp2(-\lambda_\pi) q_iq_f|a+x|-1\right]\right.  \nonumber\\
&&\left. +\Phi_0(s,m^2,\mu^2)\right\},
\label{phirho}
\end{eqnarray}
where vp means the principal value and
$m$ and $\mu$ are, respectively, the invariant masses of the final
and intermediate $\rho$ meson whose propagator is $D_\rho(\mu^2)=
m^2_\rho-\mu^2-i\mu\Gamma(\rho\to\pi\pi,\mu)$, and
\begin{eqnarray}
\Phi_0(s,m^2,\mu^2)&=&
(q_iq_f)^2\left[2a+(1-a^2)\ln\left|{a+1\over a-1}\right|\right]
\nonumber\\ &&+(q_{\pi\pi}q_f)^2\left[2b+(1-b^2)\ln\left|{b+1\over
b-1}\right|\right]. \nonumber
\end{eqnarray}
The notations in the above expressions are
\begin{eqnarray}
a&=&(\mu^2/2-E_iE_f)/q_iq_f,   \nonumber\\
b&=&m(E_i+E_f-E_\rho)/2q_{\pi\pi}q_f,
\label{ab}
\end{eqnarray}
where
\begin{eqnarray}
q_i&=&q(\sqrt{s},m_\pi,\mu)\mbox{, }E_i=E(\sqrt{s},m_\pi,\mu), \nonumber\\
q_f&=&q(\sqrt{s},m_\pi,m)\mbox{, }E_f=E(\sqrt{s},m_\pi,m), \nonumber\\
q_{\pi\pi}&=&q(m,m_\pi,m_\pi)\mbox{, }E_\rho=E(\sqrt{s},\mu,m_\pi),
\label{defin}
\end{eqnarray}
and
\begin{eqnarray}
E(M,m_1,m_2)&=&(M^2+m_1^2-m_2^2)/2M,  \nonumber\\
q(M,m_1,m_2)&=&\left\{\left[M^2-(m_1+m_2)^2\right]\right.\nonumber\\
&&\left.\times
\left[M^2-(m_1-m_2)^2\right]\right\}^{1/2}/2M
\label{energym}
\end{eqnarray}
are the expressions for energy and momentum, respectively. The decay
kinematics of the first two diagrams in Fig.~\ref{fig2}(a) result in a very
slow variation of their contribution with the change of $\lambda_\pi$.
This is
because the $\pi\pi$ cutting contributes considerably and it does not depend
on $\lambda_\pi$ (see the details in \cite{ach94a,ach94b}).
Numerically, one obtains $\Phi_{\rho\pi}(m^2_\phi,m^2_\rho)=
0.44$, 0.45, 0.47, 0.49 at $\lambda_\pi=$0, 1, 2, 4 $GeV^{-2}$, respectively.
The slight increase with $\lambda_\pi$ is due to the fact that the first two
diagrams in Fig.~\ref{fig2}(a) are opposite in sign at $\sqrt{s}<1.1$ GeV.
The third diagram in Fig.~\ref{fig2}(a), at $\sqrt{s}=m_\phi$, amounts to
$-3.4\times10^{-2}$, provided the slope of the
$\rho$ exchange is $\lambda_\rho=2\mbox{GeV}^{-2}$. The latter value is
chosen from the demand that the phase of the $\pi\pi$ scattering at this
energy range is given by the phase of the $\rho$
propagator with an accuracy
of about $10\%$. Hence, its contribution  can be neglected in comparison
with $\Phi_{\rho\pi}$.
The contribution of the diagrams in Fig.~\ref{fig2}(b) come from the $K\bar K$
intermediate states with the $K^\ast$ exchange. In the case of  $\phi$
meson it can be written as
\begin{equation}
g_{\phi\rho\pi}^{(K\bar K)}(s,m^2)=g_{\phi K\bar K}\Phi_{K\bar K}(s,m^2),
\label{imkk}
\end{equation}
where
\begin{eqnarray}
\Phi_{K\bar K}(s,m^2)&=&g_{K^{\ast+}K^+\pi^0}g_{K^{\ast+}K^+\rho^0}
\nonumber\\
&&\times{q^2_{K\bar K}\over8\pi\sqrt{s}q_{\rho\pi}}
\int_{-1}^{+1}dx{1-x^2\over a_{K\bar K}+x}     \nonumber\\
&&\times
\exp\left[2\lambda_{K^\ast}q_{K\bar K}q_{\rho\pi}(a_{K\bar K}+x)\right].
\label{phikk}
\end{eqnarray}
Here $a_{K\bar K}=(m^2_K-m^2_{K^\ast}+m^2)/2q_{K\bar K}q_{\rho\pi}$,
$q_{K\bar K}=q(\sqrt{s},m_K,m_K)$, and
$q_{\rho\pi}=q(\sqrt{s},m,m_\pi)$.
The $K\bar K$ intermediate state contribution to $g_{\omega\rho\pi}$ is
written in a similar way, with the SU(3) relation
\begin{equation}
g_{\omega K\bar K}=-g_{\phi K\bar K}/\sqrt{2}
\label{su3}
\end{equation}
being taken into account. Note also that SU(3) predicts
$g_{K^{\ast+}K^+\rho^0}=g_{\omega\rho\pi}/2$ and fixes the relative signs
of bare coupling constants in the VPP and VVP vertices.
Numerically, the effect of
$\Phi_{K\bar K}\not=0$ is negligible for $\omega$ meson because
$g_{\omega\rho\pi}^{(K\bar K)}(m^2_\phi,m_\rho^2)
/g_{\omega\rho\pi}|\simeq3\times10^{-3}$. In the case
of $\phi$ meson, at first sight this effect being expressed as the phase of
the coupling  constant $g_{\phi\rho\pi}$ is proportional to
$g_{\phi\rho\pi}^{(K\bar K)}(m^2_\phi,m_\rho^2)
/g_{\phi\rho\pi}$ and seems to be enhanced by the factor
of $g_{\omega\rho\pi}/g_{\phi\rho\pi}\simeq18$. Yet even in this case the
contribution of the $K\bar K$ intermediate state is smaller,
at $\sqrt{s}=1020\mbox{ }(1050)$ MeV,  than $6\%$ ($18\%$) of the magnitude
of the $\phi\rho\pi$ effective coupling constant.
These estimates are obtained
at $\lambda_{K^\ast}=0\mbox{ GeV}^{-2}$ and $m=m_\rho$. A more realistic
$\lambda_{K^\ast}=1\mbox{ GeV}^{-2}$,
together with
the fact that it is the averaging of $\Phi_{K\bar K}(s,m^2)$ over
$\pi\pi$ mass spectrum that enters into the expression for the $\phi\omega$
interference phase [see Eq.~(\ref{aver}) below],
both result in dividing the above estimates by the factor
of two. In the meantime,
the dominant effect of $\Phi_{\rho\pi}\not=0$ is relatively
large;
one should take into account the entire chain of rescatterings in the
diagrams of Fig.~\ref{fig2}(a). This can be made in a manner resembling the
solution of the Dyson-like equation for the vertex function. Taking the
above remarks into account, the coupling constants of $\phi$ and $\omega$
with $\rho\pi$ can be written as
\begin{eqnarray}
g^{(0)}_{\omega\rho\pi}(s,m^2)&\simeq&
\mbox{Re}g^{(0)}_{\omega\rho\pi}/\left[1-i\Phi_{\rho\pi}(s,m^2)\right],
\nonumber\\
g^{(0)}_{\phi\rho\pi}(s,m^2)&\simeq&
\mbox{Re}g^{(0)}_{\phi\rho\pi}/\left[1-i\Phi_{\rho\pi}(s,m^2)\right]
\nonumber\\
&&+ig_{\phi\rho\pi}^{(K\bar K)}(s,m^2).
\label{coupling}
\end{eqnarray}
Of course,
$\mbox{Re}g^{(0)}_{\phi(\omega)\rho\pi}$
should be determined from the partial
width of the decay $\phi(\omega)\to\pi^+\pi^-\pi^0$ on the $\phi(\omega)$
mass shell. As  is evident from Eq.~(\ref{coupling}),
the most essential contribution to the imaginary parts of coupling constants
 coming from the
$\rho\pi$ intermediate state cancels from their ratio. However,
a nonzero $\Phi_{\rho\pi}$ enters the expression for the $3\pi$ decay width
of $\omega$ and $\phi$ mesons \cite{ach94a,ach94b},
\begin{equation}
\Gamma_{\omega(\phi)3\pi}(s)=\left[\mbox{Re}g^{(0)}_{\omega(\phi)
\rho\pi}\right]^2W(s)/4\pi,
\label{3piw}
\end{equation}
where the phase space factor for the decay is
\begin{eqnarray}
W(s)&=&{1\over2\pi}
\int_{2m_\pi}^{\sqrt{s}-m_\pi}dmm^2\Gamma_{\rho\pi\pi}(m^2)
q^3_{\rho\pi}(m)      \nonumber\\
&&\times\int_{-1}^1dx(1-x^2)\left|{1\over |D_\rho(m^2)Z(m^2)}
\right.     \nonumber\\
&&\left.+{1\over |D_\rho(m^2_+)Z(m^2_+)}+{1\over |D_\rho(m^2_-)Z(m^2_-)}
\right|^2.
\label{w}
\end{eqnarray}
In the above equation, the invariant  squared masses of the charged $\rho$
mesons are
\begin{equation}
m^2_\pm=(s+3m^2_\pi-m^2)/2\pm 2xq_{\rho\pi}q_{\pi\pi}\sqrt{s}/m,
\label{mpm}
\end{equation}
with $q_{\rho\pi}=q(\sqrt{s},m,m_\pi)$, $q_{\pi\pi}=q(m,m_\pi,m_\pi)$
evaluated via Eq.~(\ref{energym}), and $Z(m^2)=1-i\Phi_{\rho\pi}(s,m^2)$.
The effect of $\Phi_{K\bar K}\not=0$ on the $\phi\to3\pi$ partial width
is negligible.

The dominant contributions to Im$\Pi_{\phi\omega}$ come from the real
$K\bar K$ and $\rho\pi$ intermediate states,
\begin{equation}
\mbox{Im}\Pi_{\phi\omega}(s)=\sqrt{s}
\left[{\mbox{Re}g^{(0)}_{\phi\rho\pi}\over\mbox{Re}g^{(0)}_{\omega\rho\pi}}
\Gamma_{\omega3\pi}(s)
-{\Gamma_{\phi K\bar K}(s)\over\sqrt{2}}\right],
\label{impi}
\end{equation}
where
\begin{equation}
\Gamma_{\phi K\bar K}(s)={g^2_{\phi K\bar K}[q(\sqrt{s},m_K,m_K)]^3
\over3\pi s}
\label{gamk}
\end{equation}
is the $K\bar K$ partial width of the $\phi$.
To gain an impression of the role of these contributions
to $\mbox{Im}\Pi_{\phi\omega}(s)$, we evaluate them
at $\sqrt{s}=m_\phi$. The $\pi^+\pi^-\pi^0$ intermediate state contribution
is, at most, $\simeq0.015\mbox{ GeV}^2$ in the model of weak $\phi\omega$
mixing and vanishes in the model of strong $\phi\omega$ mixing. The
contribution of the $K\bar K$ intermediate state amounts to $\simeq
3\times10^{-3}\mbox{ GeV}^2$. Note that the difference between the considered
models of the mixing in their predicitions for this intermediate state
is far below the accuracy (see below) of the SU(3) relation (\ref{su3})
necessary to obtain the  numbers given above.
Here we set this accuracy to be, conservatively, $20\%$.
The radiative $\pi^0\gamma$ and $\eta\gamma$
intermediate states do not exceed, respectively,  $4\%$ and $2\%$ of
the $K\bar K$ intermediate state. These figures are far below the accuracy
of SU(3) symmetry necessary to relate the couplings of the $\phi$ and
$\omega$ to $K\bar K$. Hence, the radiative intermediate states can be
neglected \cite{fn1}.

Note, for the sake of completeness, that
although the effects of $\Phi_{\rho\pi}\not=0$ are important for the
$\omega\rho$ interference pattern in the $\pi^+\pi^-$ mass spectrum
\cite{ach94a,ach94b}, in the case of the
calculation of the branching ratio of the decay to $3\pi$ they can be
modeled, at given $s$,
by inclusion of the form factor of the type
\begin{equation}
C_{\rho\pi}(s)=[1+(R_{\rho\pi}m_\omega)^2]/(1+R^2_{\rho\pi}s),
\label{c}
\end{equation}
so that  the $\omega\to\rho\pi$ vertex should now include the substitution
\begin{equation}
\mbox{Re}g_{\omega\rho\pi}\to\mbox{Re}\tilde g_{\omega\rho\pi}=
C_{\rho\pi}(s)g_{\omega\rho\pi}.
\label{c1}
\end{equation}
The effect of this substitution on the $e^+e^-\to3\pi$ cross section
behavior was discussed in Ref.~\cite{ach94a}.

\section{Evaluating the $\phi\omega$ interference phase}
\label{sec4}

The expression for the
cross section of the reaction $e^+e^-\to\pi^+\pi^-\pi^0$ that incorporates
the above features of the decay $\phi\to\pi^+\pi^-\pi^0$ can be written,
near $\sqrt{s}=m_\phi$, as
\cite{ach92,ach93}
\begin{eqnarray}
\sigma_{3\pi}(s)&=&\frac{4\pi\alpha^2W(s)}{s^{3/2}}\left|{g_{\gamma\omega}
(s)g_{\omega \rho\pi}(s)\over m^2_\omega-s-i\sqrt{s}\Gamma _\omega(s)}
\right.     \nonumber\\
&&\left.+{g_{\gamma\phi}(s)g_{\phi\rho\pi}(s)\over
m^2_\phi - s - i\sqrt{s}\Gamma _\phi(s)}\right|^2,
\label{crossse}
\end{eqnarray}
where the equations
\begin{eqnarray}
g_{\gamma\omega}(s)&=&
g^{(0)}_{\gamma\omega }-\varepsilon_{\phi\omega}(s)g^{(0)}_{\gamma\phi},
\nonumber\\
g_{\gamma\phi}(s)&=&
g^{(0)}_{\gamma\phi }+\varepsilon_{\phi\omega}(s)g^{(0)}_{\gamma\omega},
\nonumber\\
g_{\omega\rho\pi}(s)&=&
\mbox{Re}\tilde g^{(0)}_{\omega\rho\pi}(s)-
\varepsilon_{\phi\omega}(s)\mbox{Re}
g^{(0)}_{\phi\rho\pi}
\simeq\mbox{Re}\tilde g^{(0)}_{\omega\rho\pi}(s),
\nonumber\\
g_{\phi\rho\pi}(s)&\simeq&
\mbox{Re}g^{(0)}_{\phi\rho\pi}+\varepsilon_{\phi\omega}(s)\mbox{Re}
\tilde g^{(0)}_{\omega\rho\pi}(s)      \nonumber\\
&&+i\langle g_{\phi\rho\pi}^{(K\bar K)}(s)\rangle
\label{physcou}
\end{eqnarray}
relate the coupling constants of physical states whose total widths are
$\Gamma_{\phi,\omega}(s)$, with those ideally mixed.
We omit here the contribution of heavier $\omega^\prime$,
$\omega^{\prime\prime}$ resonances for the reason explained in the end
of the section. In principle, they can be incorporated in a way presented
in Ref.~\cite{ach98}. In the above formula,
$\langle g_{\phi\rho\pi}^{(K\bar K)}(s)\rangle=
g_{\phi K\bar K}\langle\Phi_{K\bar K}(s)\rangle$, and
\begin{equation}
\langle\Phi_{K\bar K}(s)\rangle=\int_{2m_\pi}^{\sqrt{s}-m_\pi}dm
\frac{2m^2\Gamma_\rho(m)}{\pi|D_\rho(m^2)|^2}
\Phi_{K\bar K}(s,m^2)
\label{aver}
\end{equation}
is the averaging over the $\pi\pi$ mass
spectrum, which corresponds to  some approximate way of taking into
account the dependence of $\Phi_{K\bar K}$ on the invariant mass.
Numerically, it reduces, at $\sqrt{s}\simeq m_\phi$,
to the diminishing of $\Phi_{K\bar K}$ by $33\%$
from its value at the $\rho$ mass.
Note that
$g^{(0)}_{\gamma V}=m^{(0)2}_V/f^{(0)}_V$ ($V=\omega, \phi$) is the
$\gamma\to V$ transition amplitude, and $f^{(0)}_V$ enters the leptonic
width of an unmixed state $V^{(0)}$ as
\begin{equation}
\Gamma(V^{(0)}\to e^+e^-,m^{(0)2}_V)=\frac{4\pi\alpha^2m^{(0)}_V}
{3f^{(0)2}_V},
\label{lepw}
\end{equation}
with
$\alpha=1/137$ being the fine structure constant. If all coupling constants
and the $\phi\omega$ mixing parameter in Eq.~(\ref{crossse}) were real, the
phase of the $\phi\omega$ interference would be given by the sign of the
ratio
\begin{equation}
R_0(s)={g_{\gamma\phi}(s)\mbox{Re}g_{\phi\rho\pi}\over
g_{\gamma\omega}(s)\mbox{Re}\tilde g_{\omega\rho\pi}(s)}.
\label{r}
\end{equation}
In the meantime, the location of the $\phi\omega$ interference minimum
in the energy behavior of the $e^+e^-\to\pi^+\pi^-\pi^0$ reaction cross
section,
\begin{equation}
s^{1/2}_{\rm min}=\left[{m^2_\phi+R_0(m^2_\phi)m^2_\omega
\over 1+R_0(m^2_\phi)}\right]^{1/2},
\label{smin}
\end{equation}
is experimentally determined to be at $s^{1/2}_{\rm min}=1.05$ GeV
\cite{cmd2,dol}.
This corresponds to $R_0=-0.13$, hence the canonical phase
$180^\circ$.  However, the above
discussion shows that considerable imaginary parts to both the coupling
constants and mixing parameter arise via unitarity, due to the real
intermediate states. As  can be observed by comparing Eqs.~(\ref{expcro})
and (\ref{crossse}) [see also Eq.~(\ref{physcou})], a sizable additional
phase $\Delta\chi_{\phi\omega}$ comes from the phase of the combination
of the coupling constants from Eq.~(\ref{physcou}),
\begin{equation}
r(s)={\mbox{Re}g^{(0)}_{\phi\rho\pi}\over
\mbox{Re}\tilde g^{(0)}_{\omega\rho\pi}(s)}
+\varepsilon_{\phi\omega}(s)+i{\langle g^{K\bar K}_{\phi\rho\pi}(s)
\rangle\over\mbox{Re}\tilde g^{(0)}_{\omega\rho\pi}(s)}.
\label{comb}
\end{equation}
The first two terms in the above equation, taken separately, are
drastically different in magnitude in the models of strong and weak
$\phi\omega$ mixing. This is because Re$g^{(0)}_{\phi\rho\pi}$
[Re$\Pi_{\phi\omega}(s)$]
vanishes in the former [latter] model. However, this dramatic difference
cancels almost completely from the sum in Eq. (\ref{comb}) that determines
the measured quantity. Indeed, one obtains, upon using Eqs. (\ref{eps}) and
(\ref{physcou}),
that
\begin{eqnarray}
r(s)&=&{\mbox{Re}g^{(0)}_{\phi\rho\pi}\over
\mbox{Re}\tilde g^{(0)}_{\omega\rho\pi}(s)}
+i{\langle g_{\phi\rho\pi}^{(K\bar K)}(s)\rangle\over
\mbox{Re}\tilde g^{(0)}_{\omega\rho\pi}(s)}-
{1\over\Delta m^{(0)2}_{\phi\omega}-i\sqrt{s}
[\Gamma^{(0)}_\phi(s)-\Gamma^{(0)}_\omega(s)]}      \nonumber\\
&&\times\left\{\mbox{Re}\Pi_{\phi\omega}(s)
+i\sqrt{s}\left[{\mbox{Re}g^{(0)}_{\phi\rho\pi}
\over\mbox{Re}\tilde g^{(0)}_{\omega\rho\pi}(s)}\Gamma^{(0)}_{\omega3\pi}(s)
-{\Gamma^{(0)}_{\phi K\bar K}(s)\over\sqrt{2}}\right]\right\}
                    \nonumber\\
&&={\Delta m^{(0)2}_{\phi\omega}\over\Delta M^2_{\phi\omega}(s)}
\left\{{\mbox{Re}g^{(0)}_{\phi\rho\pi}\over
\mbox{Re}\tilde g^{(0)}_{\omega\rho\pi}(s)}
-{\mbox{Re}\Pi_{\phi\omega}(s)\over\Delta m^{(0)2}_{\phi\omega}}
+i{\sqrt{s}\Gamma^{(0)}_{\phi K\bar K}(s)
\over\sqrt{2}\Delta m^{(0)2}_{\phi\omega}}
\right.            \nonumber\\
&&\left.-i{\sqrt{s}\over\Delta m^{(0)2}_{\phi\omega}}
{\mbox{Re}g^{(0)}_{\phi\rho\pi}\over
\mbox{Re}\tilde g^{(0)}_{\omega\rho\pi}(s)}
\left[\Gamma^{(0)}_\phi(s)-\Gamma^{(0)}_\omega(s)
+\Gamma^{(0)}_{\omega3\pi}(s)\right]
\right\}+i{\langle g_{\phi\rho\pi}^{(K\bar K)}(s)\rangle\over
\mbox{Re}\tilde g^{(0)}_{\omega\rho\pi}(s)},
\label{r1}
\end{eqnarray}
and $\Delta M^2_{\phi\omega}(s)$ is given by Eq.~(\ref{delm}).
Since the dominant $3\pi$ decay mode of the $\omega$ is cancelled from
the expression in the square parentheses of the last line of the
above equation, and the
combination of remaining $K\bar K$ and radiative decay widths appear to be
multiplied by the factor
$\mbox{Re}g^{(0)}_{\phi\rho\pi}/
\mbox{Re}\tilde g^{(0)}_{\omega\rho\pi}(s)$,  which is
either small, $\sim1/17$,
as it takes place in the model of weak $\phi\omega$
mixing, or even vanishing, as it does in the model of strong $\phi\omega$
mixing, the last term in curly brackets can be safely neglected. As a result,
the following simplified expression for  valid $r$
with a good accuracy can be written as
\begin{eqnarray}
r(s)&\simeq&{\Delta m^2_{\phi\omega}\over\Delta M^2_{\phi\omega}(s)}
\left[{\mbox{Re}g^{(0)}_{\phi\rho\pi}\over
\mbox{Re}\tilde g^{(0)}_{\omega\rho\pi}(s)}-
{\mbox{Re}\Pi_{\phi\omega}(s)\over\Delta m^2_{\phi\omega}}
\right.        \nonumber\\
&&\left.
+i{\sqrt{s}\Gamma_{\phi KK}(s)\over\sqrt{2}\Delta m^2_{\phi\omega}}
\right]+i{\langle g_{\phi\rho\pi}^{(K\bar K)}(s)\rangle\over
\mbox{Re}\tilde g^{(0)}_{\omega\rho\pi}(s)}.
\label{comb1}
\end{eqnarray}
With the accuracy of about $5\%$, the masses and widths of ideally mixed
states are replaced hereafter with those of the physical states. Note that
the combination
\begin{equation}
g(s)=\mbox{Re}g^{(0)}_{\phi\rho\pi}/
\mbox{Re}\tilde g^{(0)}_{\omega\rho\pi}(s)
-\mbox{Re}\Pi_{\phi\omega}(s)/\Delta m^2_{\phi\omega}
\label{g}
\end{equation}
standing in  the right hand side of Eq.~(\ref{comb1})
determines the branching ratio of the $\phi$ decay into
$3\pi$. Hence, its magnitude coincides
in both models of $\phi\omega$ mixing mentioned
earlier. One can obtain from the $3\pi$ branching ratios of the $\omega$
and $\phi$ at their respective mass shells that
\begin{eqnarray}
|g(m^2_\phi)|&=&c^{-1}_{\rho\pi}(m^2_\phi)
\left[{B_{\phi3\pi}\Gamma_\phi/W(m^2_\phi)\over
B_{\omega3\pi}\Gamma_\omega/W(m^2_\omega)}\right]^{1/2}
\nonumber\\
&&\simeq0.06.
\label{num3pi}
\end{eqnarray}
When obtaining this number, the dynamical phase space factors
$W(m^2_\omega)=4.5\times10^{-4}\mbox{ GeV}^3$ and
$W(m^2_\phi)=1.3\times10^{-2}\mbox{ GeV}^3$, evaluated from Eq. (\ref{w})
under the assumption of no rescattering correction [$Z(m^2)=1$, etc.], are
used and we set $R_{\rho\pi}=0$ GeV$^{-1}$ here.
Keeping $\varepsilon_{\phi\omega}(s)\not=0$ in the
transition amplitude $g_{\gamma\phi}(s)$ gives the phase shift
$\delta\chi_{\phi\omega}=1.4^\circ$,
which is below the accuracy of calculation. Hence, the calculation of
$\chi_{\phi\omega}$ is practically model independent.

First, let us give rough estimates of the phase deviation
at the $\phi$ mass. They are obtained upon neglecting the unitarity
corrections to the coupling constants of $\omega$ and $\phi$ mesons.
Then one can obtain  the above deviation as
\begin{eqnarray}
\Delta\chi_{\phi\omega}&\simeq&
\tan^{-1}\left[{m_\phi\Gamma_{\phi KK}(m^2_\phi)
\over\sqrt{2}g(m^2_\phi)\Delta m^2_{\phi\omega}}\right]
\nonumber\\
&&-\tan^{-1}{m_\phi[\Gamma_\omega(m^2_\phi)-
\Gamma_\phi(m^2_\phi)]\over\Delta m^2_{\phi\omega}}.
\label{delchi}
\end{eqnarray}
The first term in Eq.~(\ref{delchi}) gives $6^\circ\pm1^\circ$ to
$\Delta\chi_{\phi\omega}$ and the uncertainty is solely due to the $20\%$
uncertainty of the SU(3) predictions for the vector meson couplings
to $K\bar K$. We obtain these values upon inserting the Particle Data
Group entries \cite{pdg} for masses, total widths, and branching ratios,
together with the numerical value of the combination (\ref{num3pi}). The
sign of the latter (positive) is fixed in accord with the position of the
$\phi\omega$ interference minimum in the $e^+e^-\to\pi^+\pi^-\pi^0$
reaction cross section located on the right from the $\phi$ peak \cite{dol}.
The contribution of the second term is opposite in sign
to the first one and is strongly dependent on
the $\omega$ width at the $\phi$ mass, $\Gamma_\omega(m_\phi)$.
Varying $R_{\rho\pi}$ in Eq.~(\ref{c}) from 0 to 1 GeV$^{-1}$,
which corresponds to the variation of the $\omega$ width from 200 to 120
MeV, gives the second contribution  varying from
$-26^\circ$ to $-13^\circ$.
Larger values of $R_{\rho\pi}$ would
destroy the description of the data on the cross section of the reaction
$e^+e^-\to\pi^+\pi^-\pi^0$ at the energies above the $\phi(1020)$ mass.
In fact, our previous fits \cite{ach98} gave
$R_{\rho\pi}=0.8^{+0.6}_{-0.3}\mbox{ GeV}^{-1}$.

The results of more accurate numerical evaluations are  as follows.
The uncertainties  of the calculations
come from the poor knowledge of the slopes
of the form factors that enter the unitarity relations.
If one includes the $\rho\pi$ rescattering effects,
$\Phi_{\rho\pi}\not=0$, in the consideration, the variation of
$\lambda_\pi$ in the range from 0 to 4 GeV$^{-2}$ results in a small,
$0.5^\circ$ variation of the phase $\chi_{\phi\omega}$. The variation of
$\lambda_{K^\ast}$ in the same range results in the phase variation
at about
$2^\circ$. If one includes the $20\%$ uncertainty of the flavor SU(3)
predictions in Im$\Pi_{\phi\omega}(s)$, the total uncertainty amounts to
$\pm3^\circ$.  This figure is far below the current accuracy of the data,
$\Delta\chi_{\phi\omega}=\pm17^\circ$,
and is comparable with the  accuracy
expected in the future. The calculated phase depends on the form
factor (\ref{c}) that restricts the growth of the $\omega$ width with
an energy
increase. Taking into account the above uncertainty, we find
$\chi_{\phi\omega}=165^\circ\pm3^\circ$ at $R_{\rho\pi}=0\mbox{ GeV}^{-1}$
and  $\chi_{\phi\omega}=172^\circ\pm3^\circ$ at
$R_{\rho\pi}=1\mbox{ GeV}^{-1}$. The present accuracy of
the $\chi_{\phi\omega}$ measurement still admits very large bounds for
$R_{\rho\pi}$, but the future goal of the $\pm10^\circ$ accuracy of the phase
determination will permit one to put the restriction
$R_{\rho\pi}\alt2\mbox{ GeV}^{-1}$
with the perspective to give the reliable
value of this parameter upon  further improvement of the accuracy.
Second, if one does not take into account the $\rho\pi$ rescattering effect
in the $3\pi$ decay width then, including the uncertainties pointed out
above, one obtains
$\chi_{\phi\omega}=162^\circ\pm4^\circ$ at $R_{\rho\pi}=0\mbox{ GeV}^{-1}$,
and  $\chi_{\phi\omega}=170^\circ\pm4^\circ$ at
$R_{\rho\pi}=1\mbox{ GeV}^{-1}$. Unfortunately, the
difference between the predictions of the strong and weak $\phi\omega$
mixing models for $\chi_{\phi\omega}$ at the $\phi$ mass $0.6^\circ$
is too small to be measured. However, the two mixing models can be
distinguished by their predictions for the $e^+e^-\to\pi^+\pi^-\pi^0$
reaction cross section at energies near the $\phi\omega$
interference minimum \cite{ach93}. This is due to
the influence of the $K\bar K$ intermediate state on
imaginary parts of the coupling constants and the
mixing parameter which is
strongly
energy dependent. At the $\phi$ mass, its contribution is within the
uncertainties of the calculation, but it grows upon the energy increase,
so that at energies near the
interference minimum, an additional phase due to this intermediate state
could be observed \cite{ach93}.
Of course, the study of the energy behavior of $\chi_{\phi\omega}$ illustrated
by the curve in  Fig.~\ref{fig3} would be of interest.

As far as the contribution of heavier $\omega^\prime$,
$\omega^{\prime\prime}$ resonances is concerned, we neglect it here.
At the present time, this is justifiable. Indeed,
the data \cite{cmd2} give $\sigma_{\rm bg}=0.32\pm0.22$ nb
for the cross section corresponding to the amplitude $A_{\rm bg}$ in
Eq.~(\ref{expcro}) and the $\omega(782)$ tail contribution at the $\phi$
mass is $\simeq3$ nb. On the other hand, there are estimates \cite{ach98} of
the $\omega^\prime$, $\omega^{\prime\prime}$ resonance parameters which
imply  the contribution to the $3\pi$ cross section
$\sigma_{3\pi}(\omega^\prime+\omega^{\prime\prime})\simeq0.3$ nb at the
$\phi$ mass compatible with the background $\sigma_{\rm bg}$ from
\cite{cmd2}. The $\omega(782)$ tail at the same energy is estimated to be
$\simeq3$ nb. Because the data on which the work \cite{ach98} is based are
rather contradictory, it would be misleading now to include the
contribution of heavier resonances, whose parameters are extracted from these
imperfect data. Of course, the upcoming improvement of the
$\omega^\prime$, $\omega^{\prime\prime}$ resonance parameters will by no
means invalidate the present calculation of the interference phase because
their contributions can be properly taken into account in a manner similar
to Eq~(\ref{expcro}).

\section{Conclusion}
\label{sec5}

\begin{figure}
\centerline{\epsfysize=5.5in \epsfbox{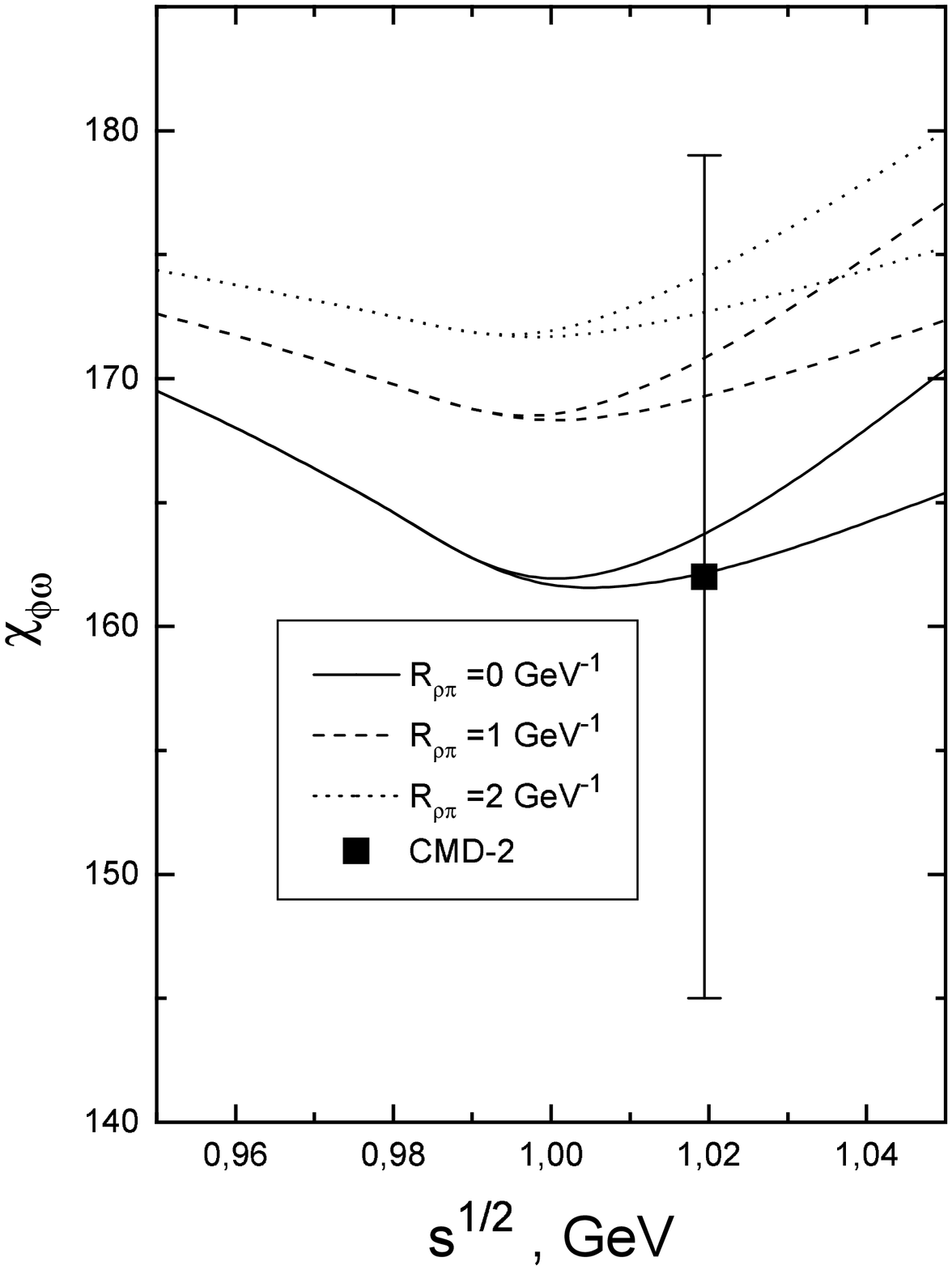}}
\caption{Energy behavior of the $\phi\omega$ interference phase in the case
of no rescattering correction to the $3\pi$ decay width, calculated at
$\lambda_{K^\ast}=1\mbox{ GeV}^{-2}$.
The splitting of each curve at $\protect\sqrt{s}\protect\geq 2m_K=0.992$ GeV
illustrates the opening of the $K\protect\bar K$
channel in the $\phi\rho\pi$ coupling (see text for explanation). The
lower curve in each pair corresponds to the latter being taken into
account.}
\label{fig3}
\end{figure}

Upon isolating possible contributions to
the $\phi\omega$ interference phase $\chi_{\phi\omega}$ in the reaction
$e^+e^-\to\pi^+\pi^-\pi^0$,
we point to the imaginary part of the $\phi\omega$
mixing parameter arising mainly due to the $\rho\pi$ state
as responsible for the deviation of
$\chi_{\phi\omega}$ from $180^\circ$ observed in the experiment
\cite{cmd2}. The uncovered source of the deviation of $\chi_{\phi\omega}$
from the naively expected phase
$180^\circ$ is far from being trivial.  The fact is that the tails of
resonances are often treated as some substitution to unknown background.
The value of information about the $\phi\omega$ interference phase obtained
in \cite{cmd2},  still to be supported by further precise measurements,
is that it give the evidence
in favor of applicability of usual field theoretical methods to
such complicated objects as hadronic resonances. The confirmation of the
observed  \cite{cmd2} deviation of the phase
would  mean that the tail of the $\omega$ is
essential at the $\phi$ mass, which is as distant from the $\omega$ as
28 widths of the latter. It can hardly be represented by the
normally used
nonresonant background. Further evidence in favor of this view could be
provided by the measurements of the energy dependence of the $\phi\omega$
interference phase as illustrated in Fig.~\ref{fig3}.
Except for
the behavior of $\chi_{\phi\omega}$, the accurate measurements of
the $\pi^+\pi^-\pi^0$ cross section in between the $\omega$ and $\phi$
peaks are necessary. They could help in an unambigous answer to the
question of the magnitude of $R_{\rho\pi}$ [Eq.~(\ref{c})], because the
cross section evaluated with  $R_{\rho\pi}=1\mbox{ GeV}^{-1}$ is lower
than
that evaluated with $R_{\rho\pi}=0\mbox{ GeV}^{-1}$ by $20\%$ ($28\%$) at
$\sqrt{s}=900$ MeV (950 MeV), and confirm a negligible role of heavier
$\omega^\prime, \omega^{\prime\prime}\cdots$ resonances
at $\sqrt{s}\alt m_\phi$.

\acknowledgments
We thank M.~N.~Achasov very much for discussion.

\begin{references}
\bibitem{cmd2}
R.~R.~Akhmetshin {\it et al.}, Phys. Lett. {\bf B434}, 426 (1998).
\bibitem{ach75}
N.~N.~Achasov, A.~A.~Kozhevnikov, and G.~N.~Shestakov, Pisma ZhETF,
{\bf21}, 497 (1975)
[JETP Lett., {\bf21  }, 229 (1975); {\bf22}, 190 (1975)(E)].
\bibitem{dol}
S.~I.~Dolinsky {\it et al.}, Phys. Rep. {\bf202}, 99 (1991).
\bibitem{ach92}
N.~N.~Achasov, A.~A.~Kozhevnikov, M.~S.~Dubrovin {\it et al.},
Int. Journ. Mod. Phys. {\bf A7}, 3187 (1992); Yad. Fiz. {\bf54}, 1097
(1991) [Sov. Journ. Nucl. Phys. {\bf54}, 664 (1991)].
\bibitem{ach93}
N.~N.~Achasov and A.~A.~Kozhevnikov, Yad. Fiz. {\bf55}, 3086 (1992)
[Sov. J. Nucl. Phys. {\bf55}, 1726 (1992)];
Particle World, {\bf3}, 125 (1993).
\bibitem{ach95}
N.~N.~Achasov and A.~A.~Kozhevnikov, Phys. Rev. D{\bf52}, 3119 (1995).
\bibitem{okubo}
S. Okubo, Phys. Lett., {\bf5}, 165 (1963).
\bibitem{zweig}
G. Zweig, CERN report 8419/TH412, (1964) (unpublished).
\bibitem{iizuka}
J. Iizuka, Progr. Theor. Phys. Suppl., {\bf 37-38}, 21
(1966).
\bibitem{arafune}
J. Arafune, M. Fukugita and Y. Oyanagy, Phys. Lett.,
{\bf B70}, 221 (1977).
\bibitem{geshken}
B.V. Geshkenbein and B.L. Ioffe, Nucl. Phys. {\bf B166}, 340 (1980).
\bibitem{svz}
M.A. Shifman, A.I. Vainshtein and V.I. Zakharov, Nucl. Phys.,
{\bf B147}, 448 (1979).
\bibitem{appel}
T. Applequist and H.D. Politzer, Phys. Rev. Lett., {\bf34},
43 (1975).
\bibitem{fn2}
The OZI rule is easy formulated at the quark level as the suppression of
the three-gluon annihilation \cite{appel}. At the level of real
intermediate states, it is formulated as the condition of compensation
of their contribution to the OZI rule violating amplitude, cf.
H.~J~Lipkin, Nucl. Phys. {\bf B291}, 720 (1987);
P.~Geiger and N.~Isgur, Phys. Rev. D{\bf44}, 799 (1991), Phys. Rev. Lett.
{\bf67}, 1066 (1991);
N.~N.~Achasov and A.~A.~Kozhevnikov, Phys. Rev. D{\bf49}, 275 (1994).
The restriction to the only $K\bar K$ intermediate state, as it was assumed
by R.~Tingen and J.~Willrodt, Nuov. Cim. {\bf 43A}, 495 (1978), could
result in some unacceptable features of the $s-$dependence of
$g_{\phi\rho\pi}$. As Tingen and Willrodt show, $g_{\phi\rho\pi}$, being
fixed to its physical value at the $\phi$ mass, grows rapidly in the
spacelike region of $s$ where it reaches  the magnitude comparable with
$g_{\omega\rho\pi}$. This behavior contradicts to the QCD qualitative
picture which predicts that the typical OZI rule violating effects are small
in this region. See \cite{svz}.
\bibitem{ach94a}
N.~N.~Achasov, A.~A.~Kozhevnikov, Int. Journ. Mod. Phys.{\bf A9},
527 (1994); Yad. Fiz. {\bf56}, 191 (1993) [Phys. Atom. Nucl. {\bf56},
1261 (1993)].
\bibitem{ach94b}
N.~N.~Achasov and A.~A.~Kozhevnikov, Phys. Rev. D{\bf49}, 5773 (1994).
\bibitem{fn1}
There were earlier attempts,
cf. J.~Pasupathy, Phys. Rev. D{\bf12}, 2929 (1975), and R.~Tegen and
J.~Willrodt, Ref.~\cite{fn2}, to take into account some unitarity effects
in the $\phi\to\rho\pi$ amplitude. These works are restricted to the
inclusion of the $K\bar K$ intermediate state only. Pasupathy  gives the
$\phi\to3\pi$ branching ratio which falls short by the factor of 4.6 as
compared to the experimental value, while the $\phi\to\rho\pi$ decay amplitude
is purely imaginary. The latter is ruled out by the data on the $\phi\omega$
interference pattern observed in the cross section of the reaction
$e^+e^-\to3\pi$.
\bibitem{pdg}
C.~Caso {\it et al.}, The Eur. Phys. J., {\bf C3}, 1 (1998).
\bibitem{ach98}
N.~N.~Achasov and A.~A.~Kozhevnikov, Phys. Rev. D{\bf57}, 4334 (1998).
\end{references}
\end{document}